\documentclass[unnumsec,webpdf,modern,large]{plain-authoring-template}%

\begin{document}

\journaltitle{arXiv}
\copyrightyear{2026}
\pubyear{2026}
\appnotes{Original Article}

\firstpage{1}

\title[Hierarchical Unsupervised Clustering for Nanobeam Diffraction]{Hierarchical Unsupervised Clustering of Nanobeam Electron Diffraction Data Using Sparse Bragg-Vector Representations}

\author[1,$\ast$]{Ian MacLaren\ORCID{0000-0002-5334-3010}}
\author[2]{Ala Al-Afeef\ORCID{0000-0003-0403-5596}}
\author[1]{Trevor Almeida\ORCID{0000-0003-4683-6279}}
\author[3]{Rantej Bali\ORCID{0000-0002-5325-4018}}
\author[1]{Shriyar Tariq\ORCID{0009-0001-6358-8089}}
\author[4, 5]{Emily Wackan\ORCID{0009-0002-6352-8550}}
\author[4, 5, 6]{Luke Daly\ORCID{0000-0002-7150-4092}}
\author[4]{Joshua F. Einsle\ORCID{0000-0001-8263-8531}}
\authormark{MacLaren et al.}

\address[1]{\orgdiv{School of Physics and Astronomy}, \orgname{University of Glasgow}, \orgaddress{\street{Glasgow}, \postcode{G12 8QQ}, \country{UK}}}
\address[2]{\orgname{Optimum Instrument Ltd}, \orgaddress{\street{London}, \postcode{WC2H 9JQ}, \country{UK}}}
\address[3]{\orgdiv{Institute of Ion Beam Physics and Materials Research}, \orgname{Helmholtz-Zentrum Dresden-Rossendorf}, \orgaddress{\postcode{01328 Dresden}, \country{Germany}}}
\address[4]{\orgdiv{School of Geographical and Earth Science}, \orgname{University of Glasgow}, \orgaddress{\street{Glasgow}, \postcode{G12 8QQ}, \country{UK}}}
\address[5]{\orgdiv{Department of Materials}, \orgname{University of Oxford}, \orgaddress{\street{Oxford}, \postcode{OX1 3PH}, \country{UK}}}
\address[6]{\orgdiv{Australian Centre for Microscopy and Microanalysis}, \orgname{University of Sydney}, \orgaddress{\street{Sydney}, \postcode{NSW 2006}, \country{Australia}}}

\corresp[$\ast$]{Corresponding author. \href{email:email-id.com}{ian.maclaren@glasgow.ac.uk}}


\abstract{If disk detection is applied to nanobeam electron diffraction datasets, then the results are effectively a list of vectors describing the position of every diffraction peak in real and reciprocal space.  This is the natural territory for the application of clustering algorithms, and they are shown to be highly effective at decomposing such datasets and automating imaging and analysis.  {In particular, this is found to work particularly well in a hierarchical clustering approach, where a first level is concerned with finding clusters of similar diffraction vectors and a second level with grouping those diffraction vector clusters that originate from the same real space area (i.e. crystal or crystallographic variant).}  Examples are shown in both polycrystalline and single crystal (with precipitates) systems.  Additionally, automated separation of amorphous or deeply nanocrystalline components is also found to be possible in an optional third level of clustering, allowing composite images of both amorphous and crystalline components in partially crystallised samples to be easily and automatically generated.  These advances promise to increase throughput in atomic structure analysis with nanobeam diffraction, and also make finding minor components much easier.  They can also serve as a preliminary step towards more detailed crystallographic or crystal size/shape distribution analysis. }


\maketitle

\section{Introduction}
Nanobeam electron diffraction (NBED) has been around for decades, and nanobeam modes were available on microscopes back in the 1980s (\cite{cowley1999electron}).  NBED became more interesting as people started getting the capability with earlier imaging detectors to collect fast enough for the acquisition of sequential sets of diffraction patterns and line-scans (\cite{BecheNanobeam, CooperNanobeam, DaultonNanobeam}).  However, it has become especially powerful in the era of 4-dimensional scanning transmission electron microscopy (4DSTEM) (\cite{Ophus4DSTEM}), in which diffraction patterns are recorded at every point in a 2D scan, resulting in a 4-dimensional datacube with dimensions $R_x,R_y,Q_x,Q_y$, where $R$ dimensions refer to real space or scan dimensions, and $Q$ dimensions refer to reciprocal space dimensions, or axes of the diffraction pattern detector/camera.  4DSTEM became practical after about 2010, as detectors became fast enough and software became available to control the acquisition of such data (\cite{DetectorsOutlook}).  This capability built on earlier ideas for mapping 1-dimensional signals such as X-ray or electron energy loss spectra into 3D datacubes (\cite{JeanguillaumeColliex}).  NBED with 4DSTEM is now an active area of electron microscopy, with innovations in techniques, software, and applications (\cite{Ophus4DSTEM}, \cite{DetectorsOutlook}).

One of the classic ways of imaging a specific feature in a microscope is dark field imaging.  Its use goes back to the 1800s in light microscopy but was very quickly implemented in transmission electron microscopy (TEM) after its invention.  The key point with dark field imaging is that the user controls which {scattered} rays contribute to the image formation using an aperture whilst excluding the primary beam, and this allows far greater selectivity over what contributes to the image than any form of bright field imaging, which includes the primary beam (and where loss of intensity from the primary beam could occur for a host of different reasons) (\cite{WilliamsandCarter}).  Imaging of specific crystals or specific areas with a specific crystallographic ordering (e.g. crystal domains within an epitaxially grown system) has long been a key capability in transmission electron microscopes.  Imaging of amorphous regions and their separation from crystalline contrast has also long been an application, by selecting part of the diffuse diffraction ring with the aperture and ensuring there is no crystalline diffraction spot also selected thereby.  However, sample bending and defects can make dark field contrast variable across a single grain, and to segment many different grains is practically difficult and requires many image acquisitions. Also, most TEM dark field imaging was performed by selecting obvious spots or placing the aperture in a position where intensity was expected from at least somewhere in the illuminated area of the sample \cite{WilliamsandCarter}.  This makes finding unexpected features harder, although there are examples in the literature of features found by moving the beam tilt and sample tilt until a contrast appears, rather than following what is obvious and already apparent - this was done, for example, to find dark field contrast from incommensurate modulations in La-doped, Zr-rich Pb(Zr,Ti)O$_3$ (\cite{RafaelPLZTdomains}).  Nevertheless, this is time-consuming and requires a lot of luck to work. 

A development once the 4DSTEM methodology was introduced was Virtual Dark Field (VDF) imaging, originally applied as part of scanned precession electron diffraction (\cite{RauchVeron_withVDF}) and also possible (but not under that name) with early versions of the Gatan Diffraction Imaging plugin for Digital Micrograph.  This has become ubiquitous in 4DSTEM and is a highly useful methodology, supported by all the main Open Source codes for working with such data, as well as by original equipment manufacturer (OEM) software.  It was quickly realized that forming {Virtual} Dark Field images with multiple apertures was useful, and could even out contrast from crystals and improve signal to noise (\cite{Gammer_VDF, Paterson_software}).

{Additionally, packages for dealing with 4DSTEM started to include disk detection for nanobeam electron diffraction data and build methods, such as automated crystal orientation mapping and strain mapping, which work with the vector based data (\cite{Savitzky_py4DSTEM, Ophus_ACOM, pyxemorientationmapping2022}).  In these methods, all the diffraction disks were detected and their parameters were kept explicitly as lists (e.g. of diffraction peak positions in reciprocal space ($Q_x,Q_y$), intensity ($I$) etc.).  It was later realized that forming images directly from the lists of diffraction peak parameters was really straightforward and conveyed further advantages.  Firstly, images are not being calculated from the raw image data (which takes a lot of memory and compute resource), but a much sparser representation just of the features of interest.  Additionally, the disk detection step concentrates on discrete disks, and not on diffuse backgrounds, so any imaging performed with this is automatically eliminating a lot of the diffuse intensity that is a feature of traditional TEM dark field or 4DSTEM VDF imaging.  Thus, \cite{MacLarenLipsettFraserOphus_DDF} introduced \textit{Digital Dark Field} imaging, in which the sparse list is interrogated with Boolean logic to determine image intensities, rather than the earlier straightforward summation of intensity within apertures in VDF, analogous to TEM dark field imaging.  This approach improved the visibility of weak features which are difficult to image using conventional TEM dark field or 4D-STEM VDF methods (\cite{DDF2_HOLZ}).}

Clustering algorithms, first developed in the fields of anthropology (\cite{kroeber1932quantitative}) and psychology (\cite{tryon1939cluster}) have long been used in applied mathematics, computer science, and signal processing and have been the subject of active developments in the connected fields of applied mathematics and computer science since at least the 1950s (in the case of k-means, \cite{perez2019k, lloyd1982least}).  In a sense, all clustering algorithms seek to perform the same task - to find groups of vectors or data points that are similar.  Most classically, a list of vectors are provided to the algorithm, distances between each vector are calculated pairwise, most commonly with the L2 norm (i.e. Pythagorean distance in whichever dimensionality is used), and then groups with small distances between them are clustered.  Alternate ways of defining ``distance" between vectors can be used to calculate the distance matrix.  This works best with a small number of dimensions or a suitable alternate distance metric between datapoints.  Many such algorithms exist, including k-means (perhaps the best known), as well as Density-Based Spatial Clustering of Applications with Noise (DBSCAN) (\cite{Ester_DBSCAN}), used here, and many more.  These came into much more regular usage in wider science with the increased visibility of machine learning as a useful contributor to other scientific fields, the increasing uptake of Python for scientific computation and visualization, and the integration of many of these algorithms and other machine learning tools into the \texttt{scikit-learn} library for Python, which has become part of the default Python install for most users.

Clustering algorithms have, of course, been used in electron microscopy previously, and especially to scanned beam techniques where machine learning is applied to the raw diffraction patterns or features/quantities/transforms derived from them (we do not consider the large field of machine learning applied in the image domain in this work).  \cite{Martineau_UML_EM} used fuzzy clustering of c-means (\cite{Bezdek_cmeans}) to sort the output from matrix factorization methods applied directly to the diffraction patterns in a data set of scanning precession electron diffraction (equivalent to nanobeam electron diffraction, with the addition of beam precession (\cite{Rauch_SPED})).  This worked well in that study when applied to twinned GaAs nanowires, in which the total number of possibilities for crystal orientation was very small (just 2).  Some of the same authors then applied DBSCAN (\cite{Ester_DBSCAN}) as implemented in \texttt{scikit-learn} (\cite{Pedregosa-Scikit-learn}) to analyse clusters in misorientations mapped by crystal orientation mapping techniques (in this case Electron Backscatter Diffraction - EBSD) on commercial purity Titanium.  Later, \cite{EnriqueAlphaBeta} used the same tools to perform analysis of mixed $\alpha-\beta$ structures in a commercial Titanium alloy.  \cite{Yoo_UMLcepstral} applied a combination of k-means clustering with matrix decomposition using either PCA (principal component analysis) or NMF (non-negative matrix factorization) to look at growth of crystals with related structures but distinct diffraction patterns from a single-crystal matrix, and also looked at whether applying \textit{cepstral} analysis (\cite{PadgettCepstrum}) to the diffraction patterns assisted this process of classification.  More recently, \cite{Vogl_Short_Range} used k-means clustering within a neural network to separate different diffraction pattern motifs relating to different short range order structures in a Si-Ge-Sn semiconductor alloy.  Similarly, \cite{KhoBridger} combined HDBSCAN {(Hierarchical Density-Based Spatial Clustering of Applications with Noise)} with a neural network to {the analysis of} combined NBED and EDX data.  Additionally, a recent study by \cite{SerinLee_clustering} used a totally different approach to clustering with a ``marching squares" algorithm (far more sequential than the traditional clustering algorithms from computing science and signal processing) applied to NBED data from a nanocrystalline gold sample, with discrimination based on image similarity metrics between diffraction patterns.  

{More recently, other clustering or classification approaches have been developed that more explicitly draw on sparse representations of diffraction data.  \cite{Savitzky_py4DSTEM} developed a Classification Approach based on Bragg Vectors with \texttt{py4DSTEM}, leveraging the image-based histogram (aka Bragg Vector Map) of diffraction vectors in the dataset but there is no attempt within to include real space position in the clustering algorithm and the clustering purely depends on reciprocal space position.  \cite{FrancisVoyles} used DBSCAN and a more hierarchical scheme and concentrated on validation of the approach with simulated and experimental data.  This method also still seems to use the loading of large volumes of raw image-based data, albeit lazily, which still requires significant storage on the computer used for the analysis (256 GB in the one example in that work) or access to a network to load the data from a server.  In the current work, we apply density based clustering with DBSCAN in a hierarchical scheme whereby the first level identifies unique diffraction peaks that appear frequently in the dataset, and the second level identifies which diffraction peaks comes from the same real space areas (i.e. the same crystal or crystal variant).  We also show that a third level can be applied in a partially amorphous sample to recover the spatial distribution of the amorphous phase.  Additionally, there is no requirement for the use of or access to the original raw data after the feature extraction, meaning that the hierarchical clustering can be done on a standard computer or laptop without a large capacity disc, and there is no need for network access once the small peak lists have been downloaded.  This is shown to be very effective in the analysis of both a partially crystallized polycrystalline thin film and in a single crystal system containing multiple families of precipitates with different orientations and structures.  The computational needs in terms of time and memory were also benchmarked.}

\section{Methods}\label{Methods}

\subsection{Samples used and their preparation}\label{Samples}

Two samples were used to demonstrate and evaluate the approaches described in this paper. Firstly, a thin film ($\sim40$ nm) of Fe$_{60}$V$_{40}$ (henceforth referred to as FeV) was deposited onto relatively thick (270 nm) SiO\textsubscript{2} grown on a (001) Si substrate at 300K. The as-deposited films possess short-range order, lacking well-defined Bragg reflections (\cite{Kataoka_FeV}). The film surface was then exposed to $2.4\times10^{15}$ cm$^{-2}$ of 25 keV Ne$^{+}$ ions, which triggered partial crystallization (\cite{Anwar_Bali_FeV}). Penetrating Ne$^{+}$ ions induce the formation of larger crystals with a BCC structure, in the form of a continuous polycrystalline layer that extends from the film surface to a certain depth within the film, in this case $\sim20$ nm, determined by the ion fluence and ion energy (\cite{Anwar_2026}).  A FIB liftout procedure was then used to make an electron transparent specimen suitable for electron microscopy. Formation of crystallized regions on the lamella sidewalls due to the FIB preparation (\cite{Anwar_HIM_2025}) were removed using 5 keV Ga$^{+}$ ions (\cite{Roeder2015}). 

Secondly, a sample was made from a piece of Imilac meteorite held by the Hunterian Museum at the University of Glasgow (GLAHM M167).  This meteorite produced a large debris field of meteorite fragments near Imilac in Chile and is a Main Group (MG) Pallasite. This family of meteorites is characterized by the mixture of silicate minerals such as olivine grains floating in an iron-nickel metallic matrix, which formed through a process of disrupted core-mantle differentiation due to an impact event (\cite{WALTE_Forming_Pallasites}). These metal phases in Imilac include kamacite (Fe-rich phase), outer tetrataenite rim (OTR, crystallographically ordered, Ni-rich phase), cloudy zone (CZ, binary phase formed from spinodal decomposition to produce islands of ordered, Fe$_{0.5}$Ni$_{0.5}$ tetrataenite within an Fe-rich matrix), schreibersite (Ni-rich phosphate, intermetallic phase), and plessite (intergrowth region of kamacite and taenite) (\cite{Goldstein_plessite}). These phases are all common to the MG Pallasite family of meteorites (\cite{YANG_PallasitePhases_20104471}). A FIB liftout was made from a region of plessite close to the CZ and thinned to electron transparency using the ThermoFisher Helios Plasma FIB at the Kelvin Nanocharacterisation Centre at the University of Glasgow using standard protocols (\cite{Schaffer_FIB_TEM}).

\subsection{Microscopy details}\label{Microscopy}
4DSTEM was performed using a JEOL ARM200F operated at 200 kV at the Kelvin Nanocharacterisation Centre at the University of Glasgow using a custom tuned nanobeam STEM mode.  The datasets were recorded using a Quantum Detectors MerlinEM system with a single Medipix 3 chip as the electron detector and with the scans defined and controlled by Gatan Digital Micrograph.  Full technical details are provided in the Supplememtal Materials.

\subsection{Data processing and machine learning}\label{Machine Learning}
The preliminary workflow from the raw data was to convert the raw binary files to \texttt{HDF5} and then detect diffraction discs with \texttt{py4DSTEM} and convert these to a \texttt{pointsarray}.  This is then a simple \texttt{NumPy} array of vectors with the following columns $(Q_x, Q_y, I, R_x,R_y,Q_r, Q_\phi)$, where the first two are the reciprocal space positions in Cartesian coordinates, $I$ is intensity, $R_x, R_y$ are real space pixel positions, and the final two $Q_r, Q_\phi$ are the reciprocal space positions rewritten in polar coordinates.  In all cases, a minimum radius was set to exclude the primary beam, as this is present everywhere and is not informative about which crystal or amorphous area is which.  ML clustering was then performed using selected columns from the \texttt{pointsarray}.  In some cases, scalings were applied to weight the relative importance of $Q$ and $R$ components in the vectors being clustered (e.g. scaling up $Q$ by a factor of 2).  Further details are provided in the Supplemental Materials.

DBSCAN requires two dataset-scale parameters: \texttt{eps}, which defines the {density threshold for to form a cluster}, and \texttt{min\_samples}, which defines the {minimum number of points} required to form a cluster {(at least these are the names given in the \texttt{scikit-learn} implementation)}. These parameters were selected separately for each dataset and clustering level because the relevant distances depend on detector sampling, scan size, camera length, and the relative scaling applied to reciprocal-space and real-space coordinates. The values used for each analysis are reported in the relevant Results sections and should be treated as dataset-specific parameters rather than universal constants. Nevertheless, similar parameter choices have worked consistently for other datasets from different samples examined in our lab and seem to be mainly a feature of average diffraction spot spacings on the detector and on the scan size (in pixels).  Thus, similar parameter choices would probably transfer for other installations with a detector about 200-300 pixels across and the first diffraction spots about 1/4 of the way from center to edge (i.e. set by camera length), and for scans with dimensions up to about 250 pixels wide or deep (or both).  It is possible that for detectors with more pixels, radically different camera lengths, or much larger numbers of pixels in a scan, that some changes to parameters may be needed.  {Similarly, for extremely large or small grains, or for large unit cell structures with extremely short reciprocal lattice distances, some changes may be needed.  }A full summary of clustering parameters used in every reported investigation in this publication is given in Table \ref{tab:Table 1} to allow easy comparison and overview, but these details are reiterated in the results with more commentary about the reasons for these choices.

\begin{table*}
    \centering
    \begin{tabular}{|c|c|c|c|c|}
        \hline
        \textbf{Which clustering?} & \textbf{\texttt{eps}} & \textbf{\texttt{min\_samples}} & \textbf{\texttt{s}} & \textbf{Details of clustering method} \\
        \hline
        L1 - FeV & 6 & 25 & 2 & L2 norm over $sQ_x,sQ_y,R_x,R_y$\\
        L2 - FeV & 2 & 2 &  & L2 norm of cluster COMs in $R_x,R_y$\\
        L3 - FeV & 1 & 20 &  & L2 norm of $Q_x, Q_y$ of unclustered points from L1\\
        L1 - plessite & 0.25 & 30 &  & L2 norm over $Q_x,Q_y$\\
        L2 - plessite & 0.65 & 2 &  & precomputed Jaccard distances between L1 cluster images \\
        \hline
    \end{tabular}
    \caption{Table 1: Parameters and clustering details for all the different clustering operations performed on the two samples.  \texttt{s} (scaling factor) is only reported for the one clustering operation where it was used.}
    \label{tab:Table 1}
\end{table*}

The resulting cluster labels from a clustering operation were then used to select rows in the \texttt{pointsarray} for formation of plots and Digital Dark Field images (\cite{MacLarenLipsettFraserOphus_DDF}, \cite{ChoudharyMacLaren}).  In some cases, multiple images were combined into a single multicolor image by colorizing individual images (according to rationales described in the main text) and then using a \textit{lighten} algorithm where the lightest pixel value in every RGB color channel is chosen. 

\begin{figure}
    \centering
    \includegraphics[width=1\linewidth]{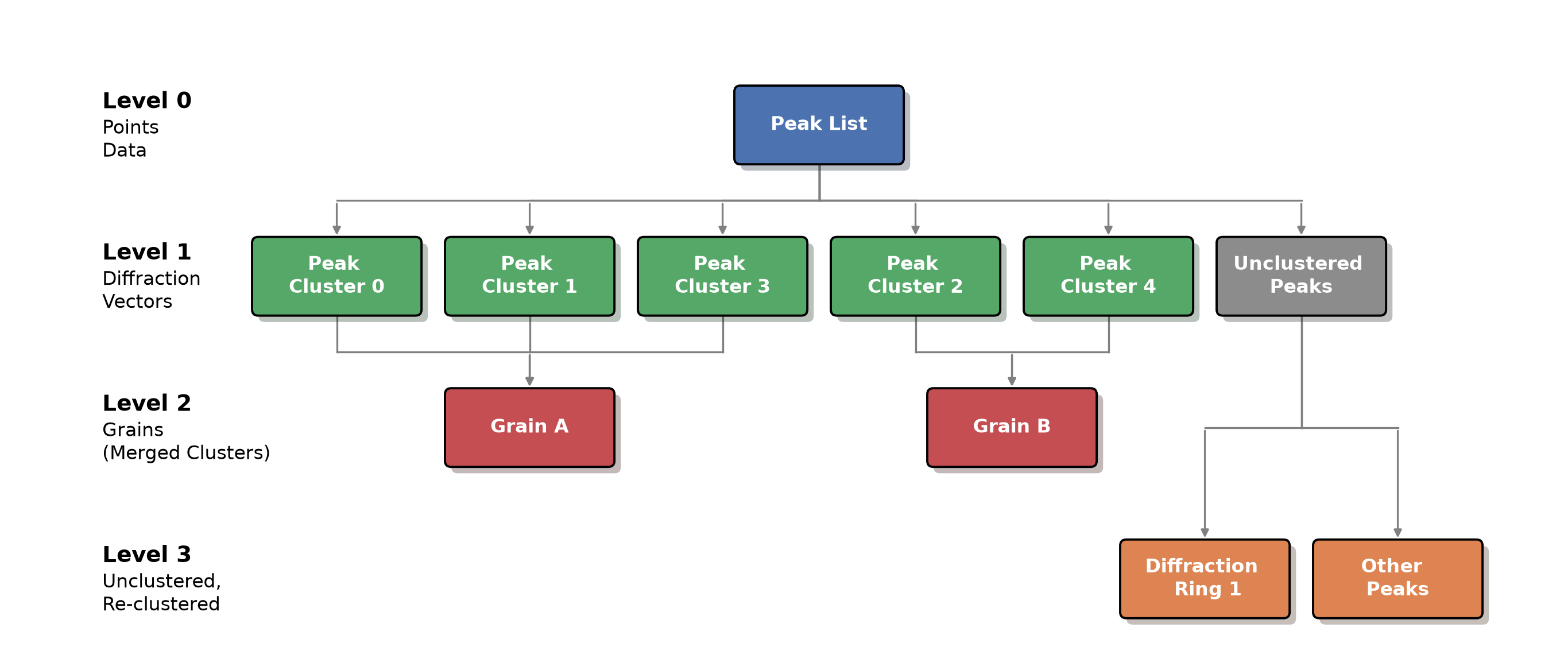}
    \caption{{A schematic flowchart of the hierarchical clustering workflow.  Level 0 is the original unclustered list of diffraction peak positions. Level 1 finds clusters of similarly oriented diffraction peaks (possibly in combination with similar positions in real space). Level 2 clusters diffraction peaks that come from the same real space locations into grain clusters. An optional Level 3 can be performed to recluster the unclustered peaks left over from Level 1, for example to find amorphous areas in the dataset.}}
    \label{fig:flowchart}
\end{figure}

For reproducibility, the clustering workflow can be summarized as follows, which is also illustrated in Fig. \ref{fig:flowchart}. First, diffraction disks were detected from the 4D-STEM datacube and exported as a \texttt{pointsarray} with one row per detected disk. Peaks within a data set-specific radius of the primary beam were excluded. {Level 1 clustering was then applied to selected \texttt{pointsarray} columns to identify individual diffraction-spot components, either from specific areas of the scan or the whole scan.  In Fig. \ref{fig:flowchart} this is illustrated for 5 clusters and some unclustered points. For polycrystalline FeV, Level 1 used scaled reciprocal-space coordinates, $[sQ_x, sQ_y, R_x, R_y]$ to identify spots from specific areas of the scan (a.k.a. crystals or grains). Level 2 clustering groups different Level 1 components that come from the same real space locations (i.e. grouping all the diffraction spots from each crystal) using some measure of spatial correlation. In Fig. \ref{fig:flowchart}, three of the L1 clusters corresponding to three diffraction peaks are connected into one grain at L2 and the other two are connected into a second grain.  An optional Level 3 clustering can be applied to Level 1 unclustered points to recover diffuse or nanocrystalline ring components from anywhere in the scan. For the plessite dataset, Level 1 clustering used reciprocal-space coordinates only, $[Q_x,Q_y]$ as similar diffraction features can come from multiple features with the same crystallography in different areas of the scan.}

\section{Results}\label{Results}

\subsection[Clustering diffraction spots from polycrystals]{Clustering diffraction spots from polycrystals together with an amorphous phase}\label{4Dclusteringofcrystals}

\begin{figure}
    \centering
    \includegraphics[width=1\linewidth]{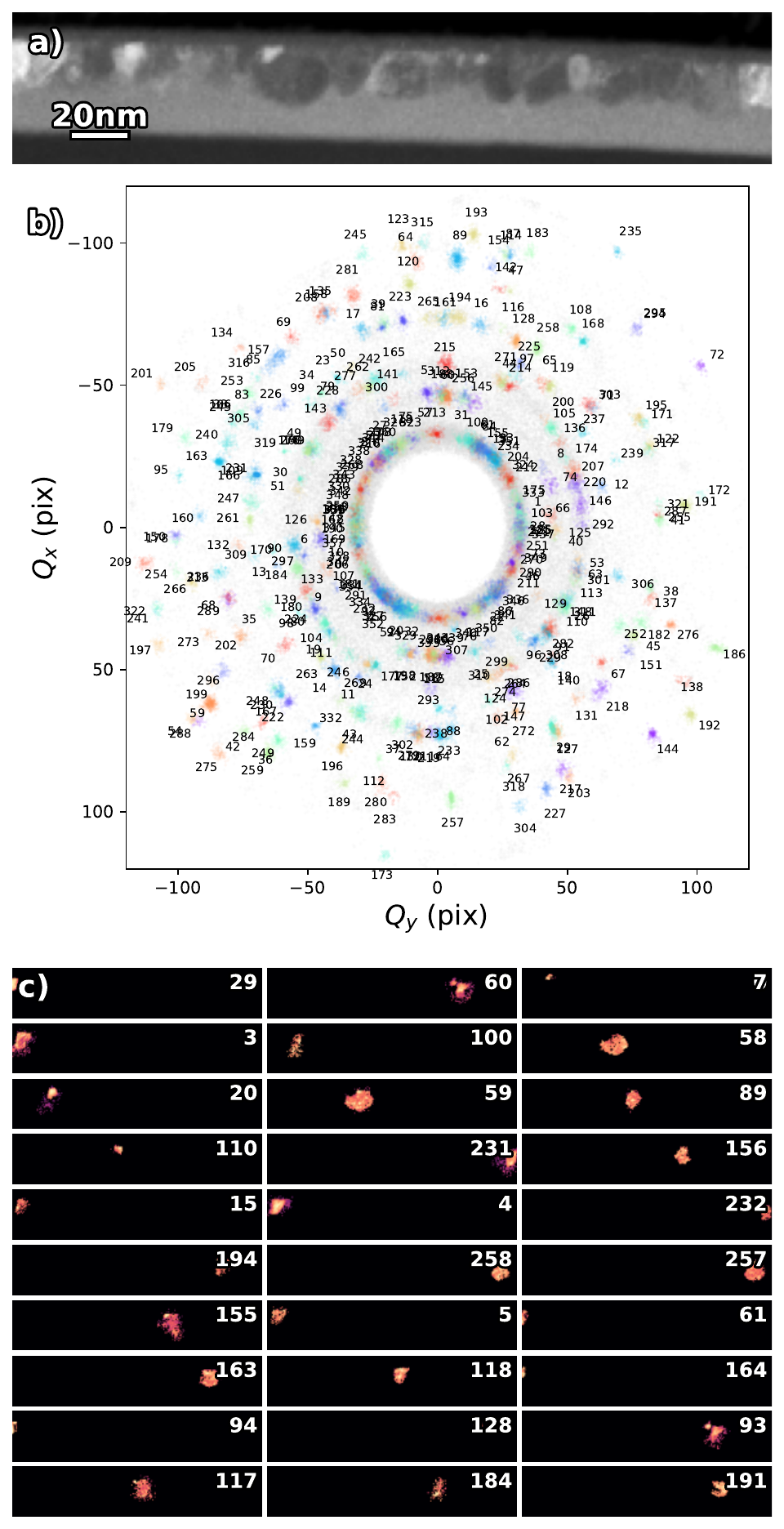}
    \caption{Level 1 clustering of diffraction spots in a thin film of FeV deposited at 300K and exposed to a Ne+ ion beam dose of $2.4\times 10^{15}$ cm$^{-2}$: a) ADF image calculated from the 4DSTEM dataset; b) the L1 cluster components plotted in terms of $Q_y$ vs $Q_x$; c) the 30 largest clusters used to form \textit{Digital Dark Field} images of their real space extent.}
    \label{fig:Fig 2_L1_FeV}
\end{figure}

For polycrystals, each diffraction spot coming from a single crystal can only originate from a restricted range of real-space positions. Additionally, if there are enough crystals there, or there is some degree of crystallographic texture, there might be similar diffraction spots (i.e. similar $Q_x, Q_y$) arising from different crystals.  Ideally, any clustering algorithm would therefore sort areas with similar $Q_x, Q_y$ values but from different $R_x, R_y$ areas in the scan into different clusters.   The obvious way to do this is to therefore cluster the array as 4-D vectors in $Q_x, Q_y, R_x, R_y$, or possibly even better, in $sQ_x, sQ_y, R_x, R_y$ so that we can scale the relative importance of real and reciprocal space in deciding what fits in which cluster.

\subsubsection[Level 1 clustering in sQx,sQy,Rx,Ry to find diffraction spots]{Level 1 clustering in $sQ_x, sQ_y, R_x, R_y$ to find diffraction spots}\label{4D L1 FeV clustering}
One ideal test object for polycrystalline clustering is a thin polycrystalline film.  Recently, \cite{Anwar_Bali_FeV} experimented with post-deposition ion-beam-induced recrystallization of initially amorphous/nanocrystalline Fe$_{60}$V$_{40}$ (henceforth known as FeV) thin films (about 40 nm thick).  The film examined in this work was deposited at room temperature and exposed to $2.4\times10^{15}$ cm$^{-2}$ Ne$^+$ ions, which was expected to lead to partial crystallization.

Fig. \ref{fig:Fig 2_L1_FeV} shows the results of applying clustering in L1 to a dataset taken from this sample. Fig. \ref{fig:Fig 2_L1_FeV}a) shows a regular Annular Dark Field (ADF) image calculated from the 4DSTEM data prior to disk detection or any detailed processing.  Fig. \ref{fig:Fig 2_L1_FeV}{b}) is a plot of Level 1 (L1) clusters against $Q_x, Q_y$. With a suitably tuned value of \texttt{eps} in DBSCAN, these can be successfully split to correspond to clusters that are consistent with visually distinct diffraction spots in reciprocal space (without merging visually distinct peaks or fragmenting individual peak clusters or making it too hard to form any clusters at all). It is recognized that not all the labels will be legible in this or any such point where hundreds of clusters are found, but the labels are shown to emphasize that all the diffraction peaks are now identified and labeled by the clustering. In this case, the following parameter choices were made:
\[\texttt{s = 2, eps = 6, min\_samples = 25}\]

\noindent The scaling parameter \texttt{s} is set to 2 to weight reciprocal space position more strongly than real space, allowing crystals to spread a little in real space, as long as the reciprocal space distribution of the diffraction spot is fairly tight.  \texttt{eps} is fairly large compared to other values used in this work (as summarized in Table \ref{tab:Table 1}), but this is because this is clustering in 4D{: the density threshold to form physically meaningful clusters scales rapidly with increasing number of dimensions}.  {The scaling factor was established by trial and error, but other scalings (e.g. 1.5) yield similar results and successful clustering is still achieved into diffraction spots coming from areas that look like grains.  However, changing \texttt{s} results in a change in density, which might mean that a change in \texttt{eps} is also needed (if \texttt{s} is reduced, a reduction in \texttt{eps} may be needed for similar selectivity).}

Fig. \ref{fig:Fig 2_L1_FeV}c) shows the largest 30 clusters (of the 359 found, plus an extra ``cluster" of unclustered points, shown in grey on Fig. \ref{fig:Fig 2_L1_FeV}b)).  {The non sequential numbering is a consequence of how DBSCAN is coded in \texttt{scikit-learn} - the clusters are labeled in order of discovery by the algorithm, not by size.}  Many of these clusters in Fig. \ref{fig:Fig 2_L1_FeV}c) clearly overlap in real space, however, such as 71, 43 and 72, which must all correspond to different diffraction spots from the same crystal.  Thus, grouping these clusters in a second level of clustering is useful in gathering unique information about individual crystals.

\subsubsection{Level 2 clustering of COMs to find crystals}\label{L2 FeV clustering}

In order to cluster the Level 1 components, we need a simple method for determining which come from the same spatial location (that is, a crystal).  The simplest method of all is to calculate the real space Center of Mass for each L1 cluster.  There are two possible definitions that could be used.  Firstly, an unweighted definition could be used where we simply count a pixel if there is any intensity at all in it in the \textit{Digital Dark Field} image:
\[COM_{\text{unweighted}} = \frac{\sum_{i=0}^{n}[{R_x}_i,{R_y}_i]}{n}\]

Alternatively, a weighted definition can be used where every pixel is weighted by intensity:
\[COM_{\text{weighted}} = \frac{\sum_{i=0}^{n}I_i[{R_x}_i,{R_y}_i]}{\sum_{i=0}^{n}I_i}\]

\noindent where \(I_i\) is the intensity of pixel \(i\) in the \textit{Digital Dark Field} image. The weighted real space COM was used here.

\begin{figure}
    \centering
    \includegraphics[width=1\linewidth]{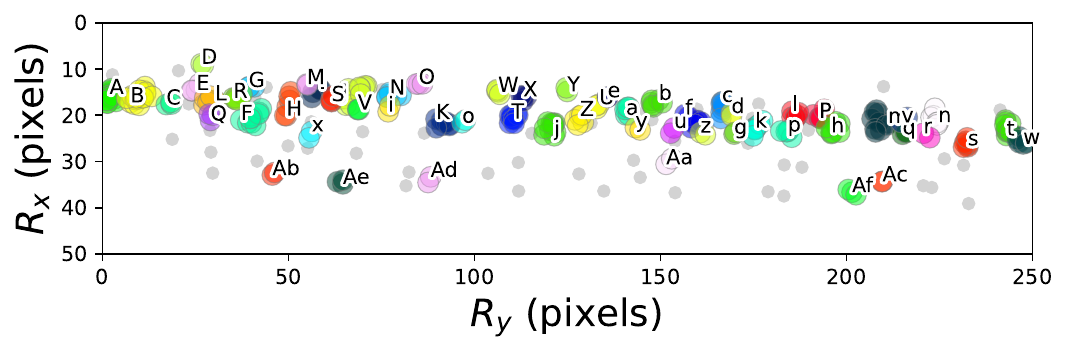}
    \caption{The COMs of L1 clusters from the partly crystallized FeV film themselves clustered into L2 clusters and labelled using letters (to avoid confusion with the numerical labels used in L1. }
    \label{fig:Fig 3 COMs clustered}
\end{figure}

\begin{figure}
    \centering
    \includegraphics[width=1\linewidth]{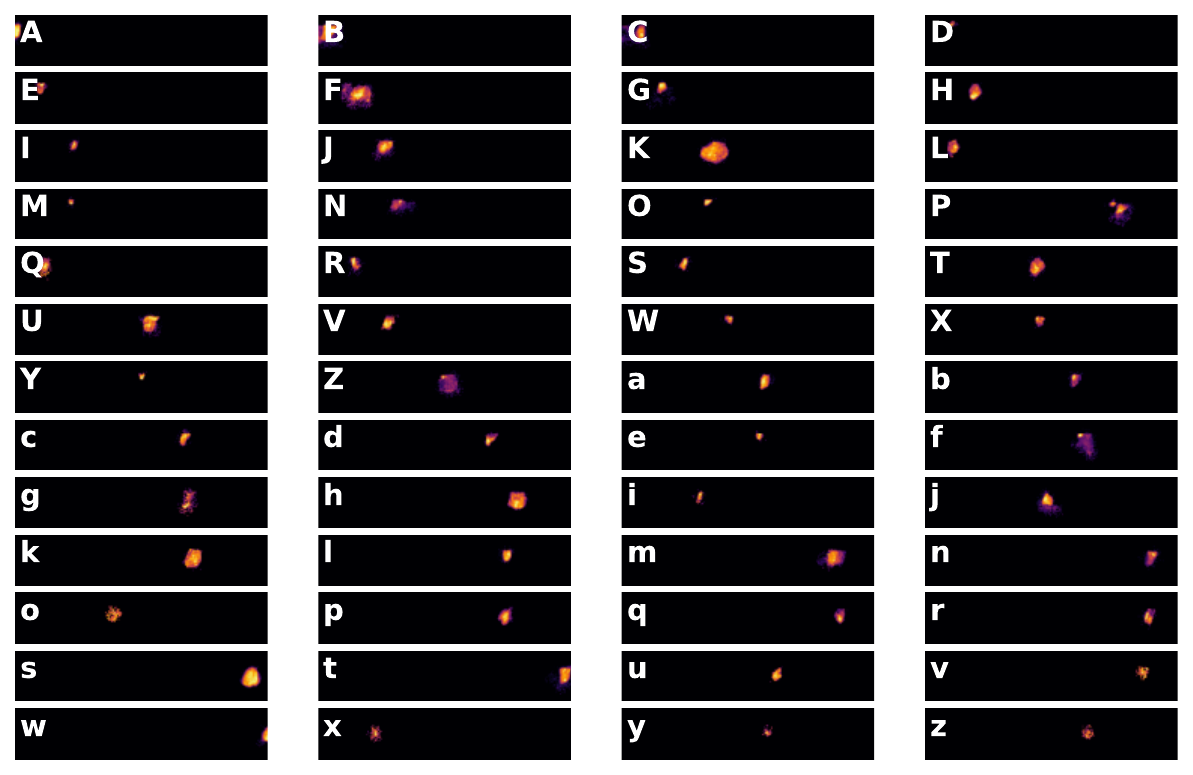}
    \caption{\textit{Digital Dark Field} Images of Level 2 clusters assigned to the crystalline region of the FeV dataset.  A display gamma of 0.5 was applied to reduce contrast differences and improve visibility of weaker regions. The images are used primarily for segmentation of crystalline components rather than quantitative interpretation of intensity. 52 clusters are found (A-Z and a-z) covering about 53\% of the film area.  A simple calculation of mean area for L2 clusters for crystals is 273 nm$^2$ giving a mean diameter of 19 nm (without performing detailed grain size measurements).}
    \label{fig:Fig 4 L2 DDFs FeV}
\end{figure}

Clustering was performed with DBSCAN and the following parameters:
\[\texttt{eps = 2, min\_samples = 2}\]

\noindent setting \texttt{min\_samples = 2}  simply means that a ``crystal" is only recognized when there are at least 2 diffraction spots coming from it.  The results of doing so are shown in Fig. \ref{fig:Fig 3 COMs clustered}.  It should be noted that the choice was made to map the cluster labels (which are numbers from \texttt{scikit-learn}) onto letters.  The results are not too sensitive to the exact setting of \texttt{eps} but setting it too high conflates nearby crystals, whereas setting it too low means that too few crystals are segmented because the criterion is too tight to allow for a little variation in COM from one diffraction spot to another in the same grain.  It is quite natural that this should occur as thickness variations and crystal bending can mean that the centers of mass for two different diffraction spots from the same grain could differ a little (this effect was already obvious many years ago if one were to make multiple dark field images from different diffraction spots from the same grain), therefore some tolerance on COMs for diffraction spots from the same grain is necessary.

Fig. \ref{fig:Fig 4 L2 DDFs FeV} shows \textit{Digital Dark Field} images of the spatial extent of the L2 clusters identified under these conditions.  These L2 clusters look very plausibly like grains in a polycrystalline material and clustering in this way is effective at separating them out.  Obviously, manual investigation of what is in each cluster can be performed to sanity-check the results.  Each cluster can also be plotted in reciprocal space to investigate further (which is shown below for a different dataset in Fig. \ref{fig:Fig8_L2_plessite}).

\subsubsection[Level 3 clustering in Qx,Qy for amorphous/nanocrystalline material]{Level 3 clustering in $Q_x, Q_y$ for amorphous/nanocrystalline material}\label{L3 FeV clustering}

However, in this particular sample, a large part of the film was expected to be amorphous or nanocrystalline and a lot of points went unclustered at L1 (101775/131337), which means either they were a tightly defined spot but from a very small crystal {below the \texttt{min\_samples} threshold, or they did not belong to something that looks like a tightly defined diffraction spot in real space.  In the specific case of samples containing amorphous materials with well-defined diffraction rings, py4DSTEM disc detection does report peaks on the ring, presumably at slight peaks in the intensity on the ring (either from noise or ``speckles" in the amorphous diffraction data, \cite{VoylesMuller}).  Thus, there should be a lot of peaks in a nanobeam diffraction dataset that connect on a ring, but don't cluster into tight spots, and which additionally will be widely distributed spatially.}  In such cases as these, it is possible to re-run clustering on the unclustered points from L1.  Two changes are made to the clustering procedure in L3 from the L1 settings.  Firstly, clustering is just done on $Q_x, Q_y$, since nanocrystalline material should display similar ring patterns wherever it occurs in the scan, and using looser definitions of what constitutes a cluster {than would be required to simply isolate single diffraction spots (e.g. 0.25 was used below for L1 clustering of $Q_x, Q_y$ positions for the plessite dataset)}:
\[\texttt{eps = 1, min\_samples = 20}\]

\begin{figure}
    \centering
    \includegraphics[width=1\linewidth]{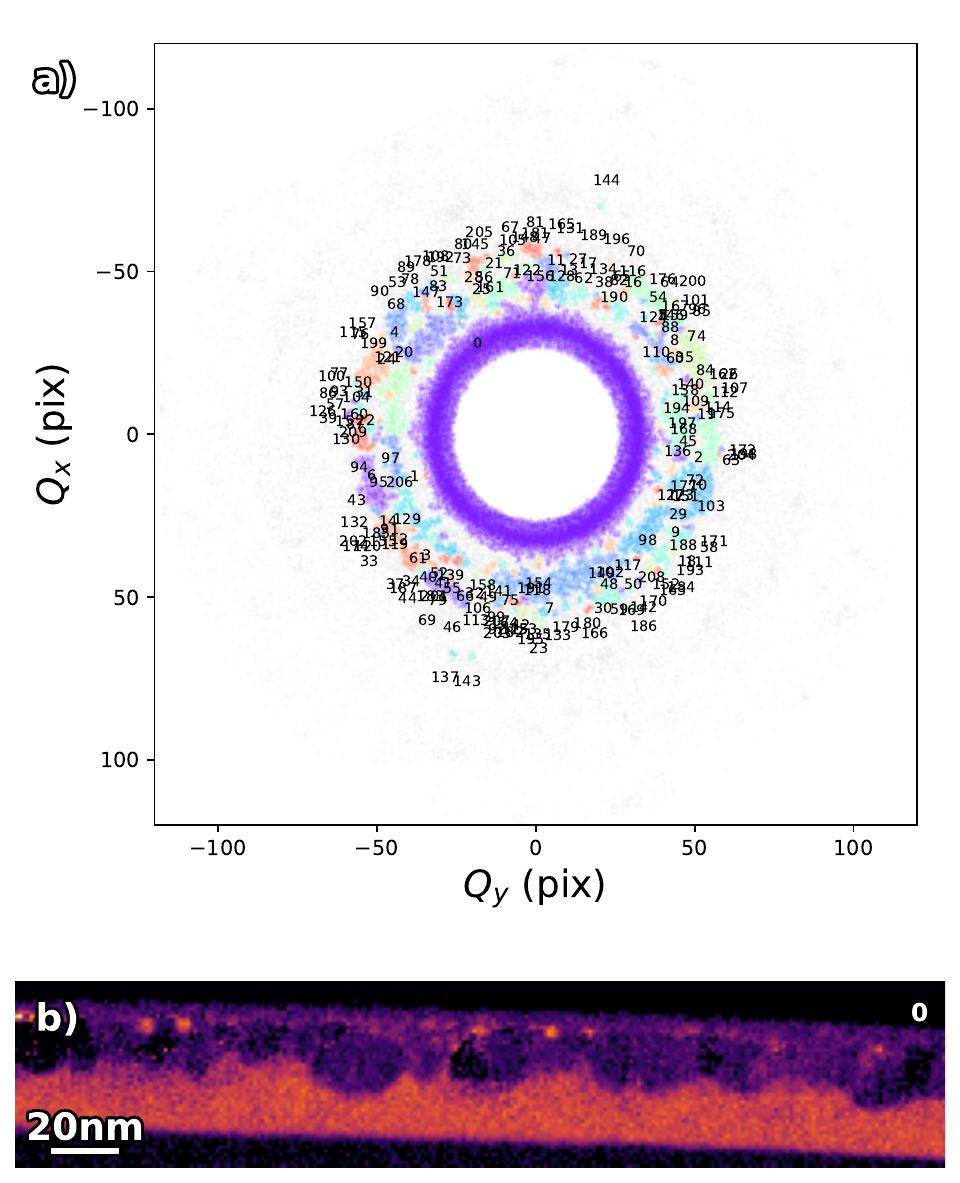}
    \caption{Level 3 clustering used to identify a diffuse/nanocrystalline ring component in the FeV dataset, corresponding primarily to the BCC FeV 110 ring: a) found by reclustering Level 1 unclustered points in $Q_x,Q_y$ using \texttt{eps = 1} and \texttt{min\_samples = 20}, the only very strong cluster is cluster 0 for the 110 ring; b) a \textit{Digital Dark Field} image made from cluster 0 with gamma = 0.5.}
    \label{fig:Fig 5 L3 nanocrystal cluster}
\end{figure}

\noindent (\texttt{eps} scales rapidly with the number of dimensions, so 1 is relatively loose for 2 dimensions, and 0.3-0.5 would find individual diffraction spots and not whole rings of spots).  Fig. \ref{fig:Fig 5 L3 nanocrystal cluster} shows the results of this L3 clustering.  Fig. \ref{fig:Fig 5 L3 nanocrystal cluster}a) shows the reciprocal space plot of clusters under these conditions and the most obvious feature is that there is one big cluster (cluster 0, plotted in purple) consisting of the entire inner diffraction ring (FeV 110), and this is the target feature, since this will be strong for any amorphous or nanocrystalline material (typically amorphous or nanocrystalline materials have a strong inner diffraction ring, even if most higher angle diffraction rings are suppressed).  Fig. \ref{fig:Fig 5 L3 nanocrystal cluster}b) shows a \textit{Digital Dark Field} image of this cluster 0 and it is clear that this image is dominated by a speckly contrast at the base of the film, below the crystals previously segmented.  There is additionally a little contrast underlying the crystals or in filaments between where they would be, suggesting some slight incomplete crystallisation.  There is also a little contrast at the top of the film, suggesting that some surface oxide was also grouped into the same cluster.  Finally, there are a few locations where features looking like very small crystals appear - probably ones that were too small for the L1 clustering parameters to detect.

\begin{figure}
    \centering
    \includegraphics[width=1\linewidth]{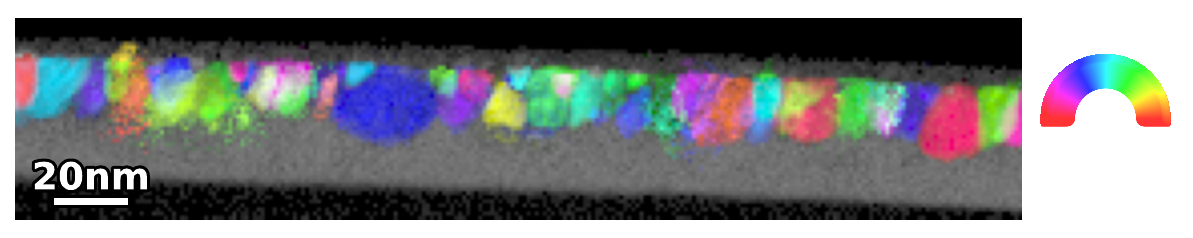}
    \caption{Composite \textit{Digital Dark Field} image from Level 2 and Level 3 clustering of the partly crystallized FeV film after ion-beam irradiation.  Crystalline Level 2 components are colored by the azimuthal angle of the brightest diffraction spot in each cluster. The Level 3 diffuse/nanocrystalline component is shown in grayscale. The color wheel indicates the mapping between azimuthal angle and displayed color.  (A version of this figure suitable for most persons with color-vision deficiency is to be found in the Supplemental Materials as Fig. S1) 
    }
    \label{fig:Fig 6 crystal / nanocrystal}
\end{figure}

Putting all this together, all crystal images from L2 clustering can be made into a color image by applying some coloring scheme to each crystal and then constructing a RGB image from the stack of images. We chose to color each L2 cluster according to the azimuthal angle of the brightest diffraction spot in that cluster. The definition for azimuthal angle is the angle measured counterclockwise from horizontal right, a definition that matches that used in, for example, Digital Micrograph and \texttt{scipy.ndimage.rotate}. The stack of color images is then transformed to a single color image by a \textit{lighten} algorithm, where the lightest value in each color channel for each real space pixel in the image stack is chosen (this is perhaps the simplest algorithm to write).  This can then be combined with the nanocrystalline FeV image from L3 clustering as grayscale, again using the \textit{lighten} algorithm.  The results are shown in Fig. \ref{fig:Fig 6 crystal / nanocrystal} and illustrate very well the partial recrystallization of the FeV film under the exposure to an ion beam.  Such segmentation and visualization is of great utility in understanding a process such as this ion-beam-induced recrystallization.  Moreover, the fact that images of all the crystals are available as a result of the L2 clustering also means that grain size and shape statistics could be calculated directly from the segmented masks, without any difficulty with overlaps, which would be of great benefit in understanding the dynamics and mechanisms of any grain growth or recrystallization process down at the submicron scale.

\subsection[Clustering for single crystal areas containing features with fixed orientation relationships]{Clustering for single crystal areas containing features with fixed orientation relationships}\label{2D Q clustering}
A simpler clustering application, at least at Level 1, is to take a single crystal or grain, possibly with subareas within with some different diffraction pattern, but with just a few possible variations in diffraction pattern related to that single crystal.  For example, this could relate to crystal lattice ordering, maybe as a result of chemical segregation (\cite{McCartan_segregation}) or domain formation, or as often occurs in epitaxial growth of oxides, where the same primitive structure persists, but there are local symmetry reductions in some layers (\cite{MacLarenLipsettFraserOphus_DDF}, \cite{DDF2_HOLZ}).  This could also result from twinning, such as in semiconductor nanowire growth, as studied by \cite{Martineau_UML_EM} with a different clustering approach.  Similarly, polymorph formation in nanowires of SiC would be another great example of this kind of thing, where the list of possibilities is few, and all with clear crystallographic relationships.  This could also occur as a result of precipitation or displacive transformations {such as $\alpha/\beta$ titanium alloys (\cite{EnriqueAlphaBeta}), shape memory alloys, or irons and steels.  The latter is} the example we will concentrate on here.  Microstructures in iron and steel alloys often contain features from the transformation between the high temperature or high alloy Face-Centred-Cubic (FCC) phase (usually known to metallurgists as austenite and to mineralogists as taenite) and the low temperature / low alloy Body-Centred-Cubic (BCC) phase (usually known to metallurgists as ferrite and to mineralogists as kamacite).  It is often found that specific orientation relationships occur relating these two phases, whatever the precise mechanisms of transformation.  Some such orientation relationships are well-known and simple to describe, such as the Kurdjumov-Sachs relationship or the Nishiyama-Wasserman relationship, although these are often found to be more of an approximation to reality than observed exactly.  

\subsubsection[Level 1 clustering in Qx,Qy to find diffraction spots]{Level 1 clustering in $Q_x,Q_y$ to find diffraction spots}\label{2D L1 clustering}
\begin{figure}
    \centering
    \includegraphics[width=1\linewidth]{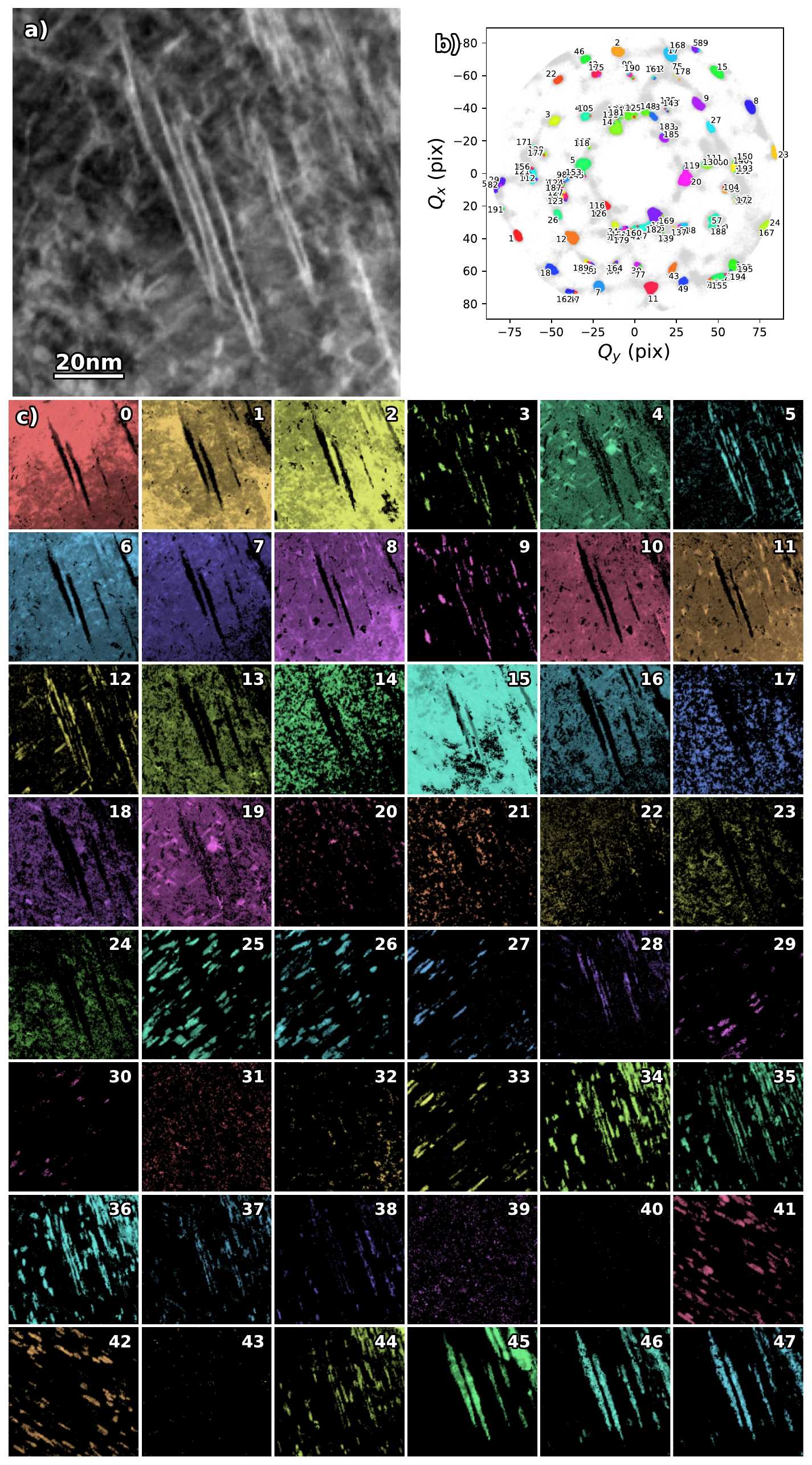}
    \caption{Level 1 clustering of diffraction spots in an area of plessite: a) ADF image of the area; b) Level 1 clusters plotted in uncalibrated reciprocal-space detector-pixel coordinates; colored points indicate clustered detections and gray points indicate unclustered detections  c) \textit{Digital Dark Field} from the first 48 Level 1 clusters, with colors to approximately match those in part b) and a gamma of 0.33 applied to flatten the contrast, displayed with gamma = 0.33.  Colors approximately match the cluster colors in b) and are assigned sequentially by cluster index and do not encode crystallographic orientation. }
    \label{fig:Fig7_L1_plessite}
\end{figure}

Fig. \ref{fig:Fig7_L1_plessite} shows the results of applying L1 clustering to this dataset using:

\[\texttt{eps = 0.25, min\_samples = 30}\]

\noindent (a relatively tight \texttt{eps} {criterion for clustering in $Q_x,Q_y$}).  

In this work, we use a scan from an area of plessite, a nanostructured mineral found in iron-nickel meteorites (\cite{Goldstein_plessite}, \cite{YANG_PallasitePhases_20104471}, \cite{NICHOLS_Plessite}) which is rich in iron and thus dominated by {body centred cubic (BCC) kamacite, $\alpha$, but does contain some precipitates of Ni rich phases (likely taenite - face-centred cubic (FCC) iron-nickel, $\gamma$, with random site occupancy by Fe and Ni, or tetrataenite, ordered FeNi, in a face-centred tetragonal (FCT) structure)} of a range of length scales.  Fig. \ref{fig:Fig7_L1_plessite}a) shows an Annular Dark Field image constructed from the dataset in the area used for the machine learning and shows a number of needle-like features (although the precise 3D morphology is unclear, and these could also be platelets).  Fig. \ref{fig:Fig7_L1_plessite}b is a plot of clusters against $Q_x, Q_y$.  With a suitably tuned value of \texttt{eps} in DBSCAN (0.25), these can be successfully split to correspond to what look like diffraction spots, neither conflating clearly distinct ones, nor splitting differences that probably don't exist or making it too hard to form any clusters at all.  As before, the images from the L1 clusters already show much more clarity about what is within the dataset.  Many clearly relate mainly to the matrix showing most of the image bright (e.g. 0,1,2,6,7,8,10,11,15).  Others relate to the larger ``needles" running from the top downwards to the right (e.g. 45,46,47).  Additional clusters show other needles in a similar orientation, but rather finer (e.g. 3,5,12,28,34,35,36,44), which were hardly to be seen in the ADF image (and not expected from the original imaging on the TEM, see Supplemental Materials Fig. S2).  There are other very fine needles pointing from top left to bottom right (e.g. 41 and 42) and from bottom left towards top right (e.g. 25,26,27,29,33, but also in some weaker clusters).  There are other clusters beyond these with different spatial variations, including some very speckly ones mainly in the matrix.  This level of nanostructural detail was far beyond anything anticipated when from standard imaging techniques conducted on this sample and clearly shows the power of this kind of ML to discover features that were not obvious in conventional contrast in STEM or TEM. 

\begin{figure}
    \centering
    \includegraphics[width=0.92\linewidth]{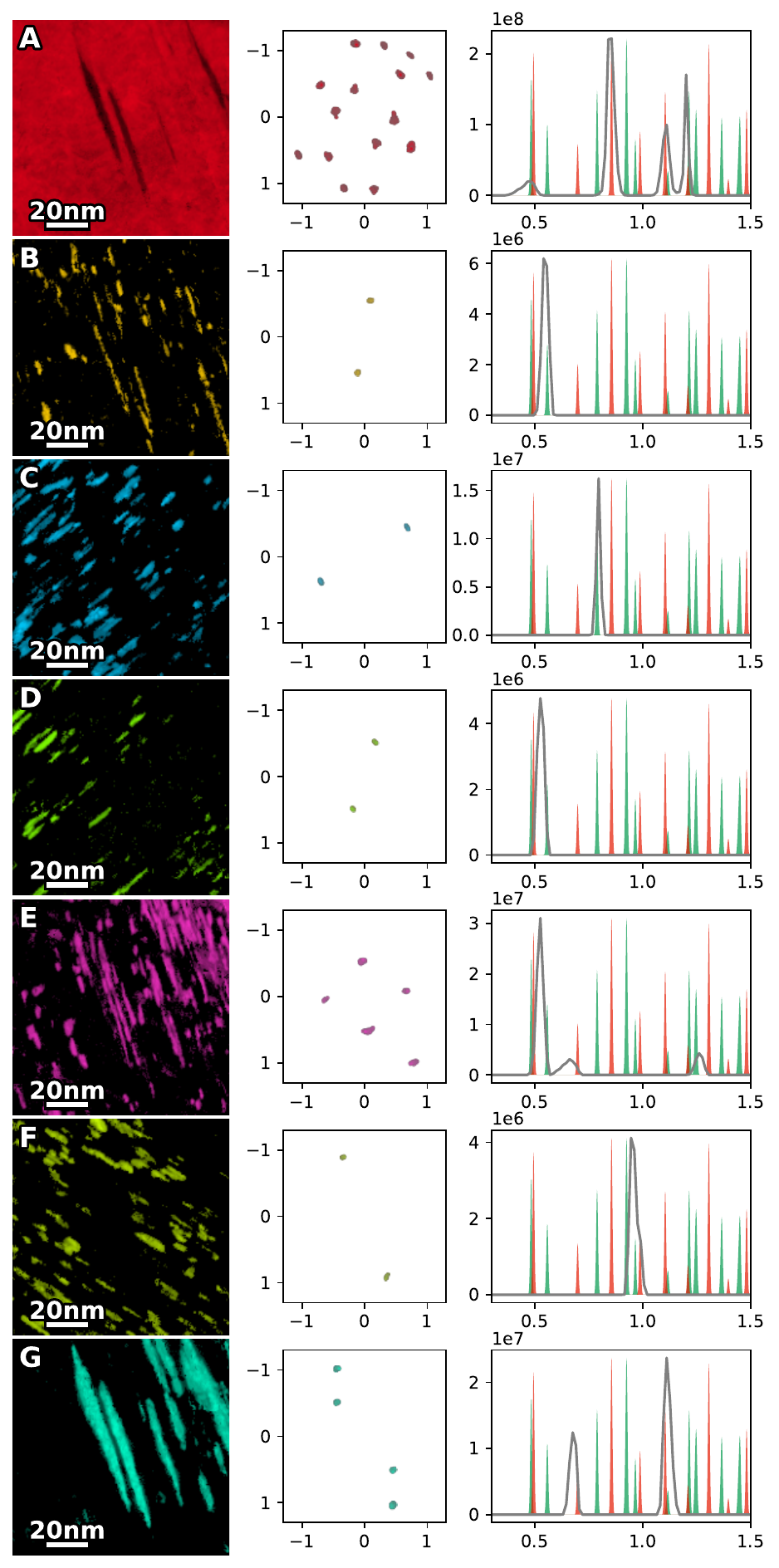}
    \caption{Level 2 clustering of diffraction spots in the plessite dataset using image-similarity grouping of Level 1 \textit{Digital Dark Field} images. Each row A-G shows, from left to right, the \textit{Digital Dark Field} image for the Level 2 cluster, the associated sparse diffraction-peak representation, and a one-dimensional radial diffractogram (both in Å$^{-1}$). Expected peak spacings for BCC Fe kamacite/ferrite (red) and FCT FeNi tetrataenite (green) are overlaid to support preliminary phase assignment. Because some residual radial distortion is present, these assignments should be interpreted as provisional until distortion-corrected virtual diffraction patterns are indexed.}
    \label{fig:Fig8_L2_plessite}
\end{figure}

\subsubsection{Level 2 clustering to find distinct families of crystals}\label{L2 plessite clustering}

{In cases where Centers of Mass are an effective discriminant between different features producing multiple diffraction spots, then all that needs to be provided to a clustering tool is the list of Centers of Mass, and the tool will use the L2 norm to calculate pairwise distances between clusters providing clear metrics with which to cluster.  }In this case of a single grain, the L1 clustering may produce a number of components all approximately centered on the center of the scan and the use of L1 Centers of Mass for the L2 clustering is unlikely to have much discriminatory power.  As such, whilst L1 clustering is easier for a single grain and only needs to consider reciprocal space vectors, more thought needs to be given to how to discriminate clusters at L2 level.  {As an alternative to simply providing a list of vectors, one can provide a 2D matrix of precomputed pairwise distances between all the features to be clustered to a clustering algorithm like DBSCAN.  For example, a precomputed distance matrix of misorientation angles was used to cluster crystal orientation data by \cite{Johnstone_DBSCAN_orix} and this approach was found to be very effective in analyzing two-phase titanium microstructures by \cite{EnriqueAlphaBeta}.  Thus we pre-compute a matrix of pairwise image ``distances" between the \textit{Digital Dark Field} images for all L1 clusters, and use this as the input for the Level 2 clustering.  In order to do so, we chose to calculate a simple IoU (Intersection over Union) / Jaccard distance metric:}
\[J(A,B) = 1-\frac{|A\cap B|}{|A\cup B|}\]

\noindent {(i.e. 1 - intersection/union).}  This ranges from 0-1 (perfect overlap $\rightarrow$ no overlap).  This is then clustered at L2 using:
\[\texttt{eps = 0.65, min\_samples = 2}\]

\noindent As before, \texttt{min\_samples}=2 means we want at least two diffraction spots to make a definite cluster corresponding to a particular crystal variant / orientation in the dataset.

The results of L2 clustering are shown in Fig. \ref{fig:Fig8_L2_plessite}.  This shows 7 distinct clusters, in each case showing a \textit{Digital Dark Field} image of the spatial extent, a plot of the cluster in reciprocal space - effectively the associated diffraction pattern, and the diffraction pattern plotted merely as a 1D diffractogram against $Q_r$.  For these diffractogram panes, the expected peak positions for BCC kamacite and FCT-tetrataenite are plotted (although FCC taenite would be similar, and the difference is in additional weak peaks for the tetrataenite).  Whilst it is apparent there is a little radial distortion in the data (probably from the tuning of the microscope optics in free lens control with no attempt to minimize radial distortion), it is possible to determine which candidate phase is most consistent with each L2 cluster.  Component A is consistent with kamacite with peaks matching the red markers for this phase, as is expected for the matrix in plessite.  B, C and F are consistent with tetrataenite/taenite with peaks matching the green markers.  D and E might be from tetrataenite/taenite but this needs further investigation.  Cluster G was unexpected.  G is clearly BCC kamacite but the diffraction pattern is very different to A so it is a different crystal orientation.  This then indicates something important about the formation of these structures, which builds on the earlier work of \cite{Goldstein_plessite} {where some tetrataenite/taenite was found to be formed by precipitation from martensite/kamacite}:
\[\alpha_2/\alpha\rightarrow\alpha+\gamma\]

\noindent where $\alpha$\textsubscript{2} refers to martensite formed on initial cooling from high temperature.  Whilst many of the clusters (e.g. C and F) are clearly consistent with this mechanism {whereby some taenite is precipitated inside the kamacite} on slow cooling, what is happening in the largest lamellae is more complex.  In the ADF image of Fig. \ref{fig:Fig7_L1_plessite}a), these larger lamellae are seen outlined by brighter material, and EELS maps of the same features find that the rims are enriched in Ni (but not the centers, as shown in Fig. S3 of the Supplemental Materials).  The clustering results show unambiguously that the rims are tetrataenite/taenite (cluster B), whereas the cores are differently oriented kamacite (G) (this is shown explicitly in the overlays shown in Fig. S4 of the Supplemental Materials).  This tells us more about the finer scale mechanisms at work in plessite, at length-scales far below those resolvable with EBSD (\cite{Goldstein_plessite}). The best explanation we can think of to generate this structural relationship is that taenite lamellae form by precipitation of a FCC Fe-Ni phase from the cooling martensite as it transforms to BCC kamacite.  The smaller lamellae then seem to remain as taenite or tetrataenite.  {But the slow cooling rates seen in these pallasite meteorites (often measured in K / million years, \cite{yang1997CZcoolingrate}) gives time for further segregation in the larger lamellae.  This results in rejection of Ni to the rims, stabilising the taenite, whilst concentrating an Fe-rich composition in the center, which eventually transforms back to kamacite.  However, this reprecipitated kamacite has a different crystal orientation from that of the matrix (presumably to reduce internal shape strain).}  As such, the transformation sequence appears to be:
\[\alpha_2/\alpha\rightarrow\alpha_{\text{matrix}}+\gamma_{\text{lamellae}}\rightarrow\alpha_{\text{matrix}}+\gamma_{\text{rims}}+\alpha_{\text{cores}}\]

To understand these processes in more detail will require more work, which would include correction of radial distortions either in microscope setup, or in postprocessing (using a standard to provide correction data, such as nanocrystalline platinum, frequently found on FIB samples), to ensure unambiguous assignment of diffraction peaks to phases.  This would then allow construction of reliable virtual diffraction patterns for each L2 cluster (capable of automated indexing).  This would then allow further exploration of which orientation relationships are consistent with the patterns with the aid of diffraction pattern simulations.  {As such, the above suggestion about the phase transition sequence should be treated as a hypothesis in need of further work to confirm it.}  Nevertheless, it is already clear that clustering on the diffraction peak position data provides a powerful tool to reveal details of complex transformations in single grains, details which were not immediately obvious from regular TEM or STEM imaging.

Additionally, it should be noted that the choice of an IoU/Jaccard metric is simple and fast to compute and performs extremely well in this case, at least in identifying areas with unique signatures in their diffraction patterns.  It should be noted that there are some weaknesses intrinsic in this distance metric for this application, and indeed the whole concept of perfect segmentation of things in a single-crystal (with ordering/precipitates) context.  Some L1 clusters relating to specific diffraction spots may belong to both the matrix and one of the precipitates, especially where they are precipitated on a specific crystallographic plane with a clear orientation relationship.  But our ``hard" clustering only allows an L1 cluster to go to one of the L2 clusters.  In this case, the common ones tend to go to the matrix cluster, the biggest available.  Moreover, Jaccard distance does not account for intensity, and just counts if there is intensity or not in defining union and intersection.  Thus, even if some precipitates are visible in an image as brighter features with a background intensity from the matrix, Jaccard will not distinguish those brighter features and treat this as indistinguishable from the matrix.  Thus, using a more sophisticated distance metric calculation may have advantages.  However, having tried another simple-to-implement metric that does include intensity, cosine distance (as used by \cite{SerinLee_clustering}), this performs far worse and is biased by the much stronger intensity in the matrix compared to everything else.  The Jaccard metric was therefore used here as a fast and interpretable measure of spatial mask overlap, rather than as a complete measure of diffraction-pattern similarity. Intensity-aware alternatives, such as weighted Jaccard or Dice/F1 similarity, may be useful in datasets where matrix and precipitate reflections overlap strongly.  Improving the image-similarity metric is therefore an important direction for future work.  However, it can also be noted that manual examination of cluster components could also be a very fruitful approach for the assembly of more detailed cluster models.  Also, using the L2 cluster images as masks to select pixels in the original \texttt{pointsarray} could provide a way to get a complete set of diffraction spots for every identified feature in the L2 cluster list.

\section{General Discussion}

Here we have demonstrated that clustering either in 2D (clustering $[Q_x,Q_y]$ vectors) or 4D (clustering $[sQ_x,sQ_y,R_x,R_y]$ vectors) on arrays of diffraction peak data is a powerful approach to make sense of nanobeam diffraction data and easily and quickly reveals {features of scientific or technological significance, once the initial step of detecting the diffraction peak positions has been performed.  In the following, we will consider a number of wider issues concerning this work, its relation to prior work, and the strengths and weaknesses of the approach.}

\subsection{Comparison to Whole Pattern Approaches}
In comparison to whole pattern approaches which work with full diffraction frames and their image correlations, this method is much faster to compute and tune and requires much less memory (see below for some computing performance metrics).  It has the advantage and disadvantage that it only works on crystalline spots that can be reliably detected.  This has the advantage of filtering out diffuse scattering (whether from inelastic scattering or other mechanisms) and will hardly be disturbed by noise in the data (even if some is falsely detected as diffraction peaks, as these will not be spatially correlated).  But this means that applications to localized diffuse scattering (e.g. from short range order) or broader diffraction peaks from glasses could be difficult and other approaches would be better.  Alternatively, other approaches to vectorizing a dataset could be explored.

\subsection{The Reliance on Disc Detection}
{It is worth noting that this technique will only work well if the disc detection step is well optimized for the dataset, which is a problem common to all methods that rely on disc detection, such as automated crystal orientation mapping (\cite{pyxemorientationmapping2022,Ophus_ACOM}), strain measurement (\cite{Savitzky_py4DSTEM}) and digital dark field imaging (\cite{MacLarenLipsettFraserOphus_DDF,ChoudharyMacLaren}).  As a result, this technique could work even better with scanning precession electron diffraction data (\cite{Rauch_SPED}), which is likely to be easier to interpret as the intensities will be more predictable and pseudo-kinematical and the spots rounder with a more even intensity profile.  Similar benefits would apply by using it on data derived from multi-angle precession electron diffraction (\cite{Ribet_MAPED}). Improved disc detection using tricks like patterned apertures (\cite{Zeltmann_Bullseye}), or neural network assistance (\cite{MunshiDeepLearning}) may improve the data quality sampled by the machine learning tools and improve the quality of the analysis results.  Nevertheless, even noisier data than that presented in this work, with a lot of spurious disk detections still produces sensible and consistent clustering outputs, simply because noise and spurious detections do not cluster spatially or in reciprocal space and just end up in the unclustered points at Level 1.}

\subsection{Parameter Selection}
{The effects of incorrect parameter selection for clustering should also be considered.  If \texttt{eps} is set too low for the dataset, then very few and rather small clusters will be found and most points will go unclustered.  Some clusters may also be split that should also be a single cluster.  Examination of the reciprocal space plot and the real space images may help diagnose this.  Similarly, if \texttt{eps} is set too large, then clusters in reciprocal space will no longer look like rounded diffraction spots and may connect spots into arcs, streaks or other irregular shapes.  However, in the experience of the authors, once a reasonable tuning has been achieved, it seems quite stable over a range of similar datasets.  We have occasionally seen the same grain split into two L2 clusters, especially where the grain is larger and shows some bending such that there is a significant change in diffraction spot intensities from one end to another, but such clusters could be grouped manually by the user, if needed, after the automated clustering.}

\subsection{Interpretation of the Results and Applications of this Approach}
Performing clustering in two or three hierarchical layers produces physically meaningful components.  In this work, segmentation of individual crystals and the residual uncrystallized nanocrystalline material was shown to be one such application.  Further analysis could then yield the size and shape statistics of the crystals and the area fraction of the recrystallized material.  The plessite example illustrates how clustering can reveal nanoscale details of diffusion-driven phase transformations in a natural Fe-Ni mineral frequently found in meteorites {and to illustrate potential phase transformation mechanisms or provide the data for further crystallographic investigation}.  It would equally be possible  these techniques to more effectively use nanobeam electron diffraction to study processes in materials, film and multilayer systems, devices, crystalline pharmaceuticals, as well as naturally occurring minerals from either terrestrial or planetary sources.  Some areas for application are outlined in the following paragraph.

One simple application would be to segment crystals (probably in 4D) and study phase transformations in \textit{in-situ} 4DSTEM experiments (e.g. studying changes as a function of time, temperature, bias [especially in electrochemical experiments]) {because the speed with which the clustering can be done allows rapid comparison of multiple datasets from different samples}.  Going beyond this, if the data quality is good enough and covers enough range to get quite a few diffraction spots per crystal, then it could be used to drive an approximate orientation mapping per crystal, as was recently done in a different clustering approach (\cite{SerinLee_clustering}).  Another simple one to implement would be to segment crystals by orientation on a given diffraction ring (which could use either $[sQ_x,sQ_y,R_x,R_y]$ or $[rQ_r,pQ_\phi,R_x,R_y]$), which could be used for preferred orientation measurement - using polar coordinates is likely quite helpful in many systems.  Another may be distinguishing two or more nanocrystalline or amorphous structures, mainly by concentrating on $Q_r$ as there will be a difference in peak positions for the two structures.  In short, if the dataset and the problem to be addressed can be written in vectors in either a cartesian or polar basis, and there is a noticeable distinction expected, then clustering should find it, after some tuning of the clustering parameters.

\subsection{Performance Benchmarking}
The other advantage of using such techniques is a practical one of major benefit to the experimenter.  Once a notebook is set up and the parameters are about right, it is relatively fast and unchallenging on memory (runs in $\sim10$s on a recent model Macbook Air laptop to go from data loading through to L1 and L2 clustering and displaying the results, and peak memory usage less than 4GB [in our example 3712 MiB = 3892 MB], see Table \ref{tab:Table2}) to use it to produce a quick analysis of what is in the dataset.  Presumably, processing times would be even faster on a dedicated inline server with a more powerful CPU.  As such, it is of great value in screening data quickly to determine if it contains features of interest, something that could be implemented in a processing computer attached to a microscope allowing fast feedback on experiments whilst the user is working.  It may also, as with the plessite example, reveal unexpected features that can then inspire and aid with the planning of more detailed experiments.  It should also be noted that whilst the initial processing of the data to detect the diffraction peak positions and save them as a \texttt{pointsarray} is relatively slow and benefits from a high specification computer (a dedicated server was used in this work), the \texttt{pointsarray} is small in memory size ($\sim50$ MB for the plessite dataset compared to the raw 4DSTEM data at $\sim 5$ GB) and easy to process on a regular desktop or laptop {(c.f. \cite{Savitzky_py4DSTEM}, who noted similar advantages for sparse list-based representations for other uses).}

\subsection{Algorithm Choice and Implementation on Different Systems}
In this work, we have just used DBSCAN (\cite{Ester_DBSCAN}), which worked well for the cases studied so far, and the references to \texttt{eps} are specific to this particular tool.  k-means has also been experimented with but is somewhat restricted in application by the fact that it requires the user to tell it how many clusters to create, and it was not felt in these cases that it was obvious how many would be appropriate.  An algorithm, such as DBSCAN, that calculates the number of clusters based on the data seems more generally useful in semi-automated analyses to provide rapid feedback and initial analysis.  Additionally, false detections of diffraction peaks are likely in processing of scanned diffraction data, which means that noise points will be present in the data and should be left unclustered by the algorithm.  However, there are many other clustering algorithms available and it is quite likely that further experimentation could find that some have particular advantages in some cases.  One point on this, however, is about speed versus capability - DBSCAN is rather simple and fast (about 11s to do all the operations up to L2 clustering on a normal laptop, see Table \ref{tab:Table2}), having tried HDBSCAN and OPTICS (as implemented in \texttt{scikit-learn} (\cite{Pedregosa-Scikit-learn})), these are more complex and slower.  So, the choice of algorithm may also depend on how fast the analysis is required to be and how much processing power and memory is available.

{Up to this point, we have just considered the application of these algorithms in scikit-learn, which operates simply in the \texttt{NumPy} ecosystem and is designed for CPU use on standard computers.  Much of machine learning, however, now runs on GPUs using other ecosystems that leverage their parallel processing strengths.  It would be possible in future to use the same approaches written for a more GPU-native ecosystem such as \texttt{PyTorch} and thereby allow extension to larger and more complex datasets and integration with more complex Machine Learning workflows.}

\begin{table}
    \centering
    \begin{tabular}{|c|c|c|}
        \hline
        \textbf{Operation} & \textbf{Time} & \textbf{Peak Memory}\\
        \hline
        Data loading and initial calibration & 2.77 s & 1585 MiB\\
        Cropping and annular DDF images & 0.46 s & 1291 MiB\\
        L1 clustering & 6.22 s & 3712 MiB\\
        Jaccard distance computation & 1.27 s & 2348 MiB\\
        L2 clustering & 0.45 s & 1444 MiB\\
        \hline
    \end{tabular}
    \caption{Timing and memory usage of key operations on the plessite dataset (Macbook Air M4, 2025).  Average of 7 iterations.} 
    \label{tab:Table2}
\end{table}

\section{Conclusions}
Clustering algorithms have been used to analyse nanobeam electron diffraction 4DSTEM datasets, which had already been preprocessed to a list of position- and intensity-tagged diffraction peak positions.  Because these are just lists of vectors, cluster analysis is extremely effective, with some dataset-specific parameter selection of clustering parameters according to the details of the dataset (especially detector and scan size in pixels).  An initial Level 1 clustering decomposes a dataset into L1 clusters relating to specific diffraction vectors from specific areas of the scan.  Level 2 clustering can then be applied to group together those L1 clusters that come from the same spatial location (e.g. grains, specific crystallographic variants or domains).  In cases of amorphous/crystalline mixes, L3 clustering can be used to find the amorphous areas.  This has been shown to work in both a partially crystallized thin film containing a significant proportion of amorphous material, and in a scan of a single crystal, within which there are smaller precipitates with fixed orientation relationships with the matrix.  Additionally, further benefits arise compared to other imaging methods.  For the polycrystalline case, images of each individual crystal can be formed, which could easily be leveraged for crystal size and shape analysis, with no difficulties of segmentation.  Also in-plane orientation of diffraction vectors is part of the results of the clustering so mapping of preferred crystal orientations would be straightforward.  In the single-crystal case, features were revealed that were not apparent by regular TEM or STEM imaging, and not detected in cursory EELS scans.  This data-driven approach will remove some user biases and highlight unanticipated results. Additionally, this type of ML application will also streamline analysis of nanobeam diffraction datasets, both in providing quick feedback on the features within, and in providing the detailed statistical and crystallographic information that could be useful in more in-depth analyses.

\section{Competing interests}
No competing interest is declared.

\section{Author contributions statement}

I.M. conceived and led the study, acquired the data on the microscope and led the development of the ML analysis.  A.A.A. assisted with ML analysis and image similarity computations.  T.A. assisted with microscopy and background on the FeV samples.  R.B. provided the FeV samples and assisted with providing context about the samples, their manufacturing, properties and purpose.  E.W. and S.T. prepared the plessite sample and E.W. additionally assisted with STEM analysis thereon.  E.W., L.D. and J.F.E. assisted with understanding the context of plessite and the wider set of micro/nano-structures observed in Fe-Ni regions in meteoritic samples.  All authors contributed to the writing of the manuscript.

\section{Acknowledgments}
RB acknowledges DFG Project no. 554722570. We thank René Hübner for preparation of FeV lamellae for TEM.  Ion-irradiation was performed at the Ion Beam Centre facility of the Helmholtz-Zentrum Dresden–Rossendorf. TA acknowledges support from EPSRC (EP/X025632/1) for the microscopy performed on the FeV sample.  LD acknowledges support from UKRI STFC grants (ST/Y004817/1, ST/T002328/1, and ST/W001128/1). The FIB used in the preparation of the plessite lamellae was provided by the EPSRC (EP/P001483/1). Colin How, Dr Sam McFadzean and Dr Kayla Fallon are thanked for helpful advice and discussions about the operation of the JEOL ARM200F and acquisition of data with the MerlinEM detector and readout system and support of this microscope facility.  IM acknowledges helpful discussions with Dr Serin Lee and Prof Colin Ophus at Stanford University, which helped to inspire the ideas that led to this manuscript.  Alicija Hertmanowska is gratefully acknowledged for experimenting with very rough early versions of these ideas on different datasets as part of an undergraduate project course and showing that the approach had promise. 

\section{Preprint Deposit}

A preprint of this article has been deposited to MacLaren \textit{et al.} (2026) (https://arxiv.org/abs/2606.23201)

\section{Data and software availability}
The clustering workflow used in this study was implemented using \texttt{Python}, \texttt{py4DSTEM}, \texttt{scikit-learn}, and functions from the \href{https://github.com/maclariz/Kelvin_STEM}{\texttt{\textcolor{blue}{Kelvin\_STEM}}} repository. Code, notebooks, and processed data are available at \href{https://researchdata.gla.ac.uk/2439/}{\textcolor{blue}{Enlighten Research Data}} detailing everything required to reproduce the analyses.  An online tutorial version of the FeV clustering notebook is available at \href{https://colab.research.google.com/drive/1hKaw7YiaZiXWbvSQ1oKRuM_84tz7xUma}{\textcolor{blue}{Google CoLaboratory}}.

\bibliographystyle{plainnat}
\bibliography{reference}
\end{document}


\title{High Throughput Analysis of Nanobeam Electron Diffraction Datasets using Unsupervised Clustering - Supplemental Materials}

\author[1]{Ian MacLaren}
\author[2]{Ala Al-Afeef}
\author[1]{Trevor Almeida}
\author[3]{Rantej Bali}
\author[1]{Shriyar Tariq}
\author[4, 5]{Emily Wackan}
\author[4, 5, 6]{Luke Daly}
\author[4]{Joshua F. Einsle}

\affil[1]{School of Physics and Astronomy, University of Glasgow, Glasgow, G12 8QQ,UK}
\affil[2]{Optimum Instrument Ltd, London, WC2H 9JQ, UK}
\affil[3]{Institute of Ion Beam Physics and Materials Research, Helmholtz-Zentrum Dresden-Rossendorf, 01328 Dresden, Germany}
\affil[4]{School of Geographical and Earth Science, University of Glasgow, Glasgow, G12 8QQ, UK}
\affil[5]{Department of Materials, University of Oxford, Oxford, OX1 3PH, UK}
\affil[6]{Australian Centre for Microscopy and Microanalysis, University of Sydney, Sydney, NSW 2006, Australia}

\maketitle

\section{Introduction}

This document presents some additional information to assist the reader in the understanding of the main paper.  This mainly consists of optional extra figures and technical details.

\section{Full details of Methods}
\subsection{Microscopy}
The data was recorded on the ARM200F using a small beam current selected by choosing a high numbered Condenser Lens setting in STEM mode (principally 9C) for this microscope.  The smallest available condenser aperture (10 \textmu m) was used to get the lowest convergence angle possible.  The probe was then manually tuned in ``Free Lens Control" to reduce the convergence angle further by reducing the excitation of the CL3 lens to approximately B000 and refocusing with the CM lens.  Some adjustment of the beam tilt and astigmatism was needed using the Ronchigram as a guide in order to produce a round probe.  Displacement of the probe on scanning was extreme after conducting this procedure, but was minimized by making small adjustments to the IL1 lens excitation.

Diffraction pattern datasets were collected at a camera length of 80cm using a MerlinEM detector and readout system with $256\times256$ pixels.  Scans were controlled by using Gatan Digital Micrograph to define the scan area.  Firstly, a standard ADF detector was used with a normal dwell time of ~100 \textmu s to set the scan area and check focus.  Then the ADF was removed, the Merlin detector inserted, and the dataset recorded with the Merlin acquisitions triggered by hardware clock pulses from the Digiscan unit.  Typically, a pixel dwell time of 6 ms was used, with a pixel exposure time of 5 ms (plus 822 \textmu s readout time) in continuous mode and recording the number of scan pixels required, plus an extra one per line for flyback.

Datasets collected under these Free Lens Control conditions do not have the same spatial calibrations as those taken under standard STEM alignments and also show some image rotation, but a pixel size calibration can be recovered by reference to an image of the same area taken under standard STEM conditions where calibrations are well known.

\subsection{Data handling and processing}
Many of the unique functions used in this work are built upon prior work but packaged into functions in the \texttt{Kelvin\_STEM} repository on GitHub.  The data from the Merlin system came as \texttt{.mib} files, which are simple binary files, which in our case were 1 byte per diffraction pattern pixel plus 384 bytes of header in each frame.  A simple function was used to read them as a \texttt{NumPy} memory map and these were then read into \texttt{py4DSTEM} and saved as \texttt{HDF5} files.  However, it was found that the diffraction patterns were flipped compared to those on the screen or other cameras on the microscope, so a flip on axis 2 of the ``datacube" was applied (the vertical direction in the diffraction patterns) before doing anything else.  An ADF image was calculated and the diffraction disks were detected, pattern centres were found, ellipticity in the diffraction patterns (which was very small on this microscope) was corrected, and the lists of Bragg peaks converted to a \texttt{pointsarray} for easy access to all the data in a form that can readily be treated as vectors - in all cases, the seven column version was used $(Q_x, Q_y, I, R_x,R_y,Q_r, Q_\phi)$, where the first two are the reciprocal space positions in Cartesian coordinates, $I$ is intensity, $R_x, R_y$ are real space pixel positions, and the final two $Q_r, Q_\phi$ are the reciprocal space positions rewritten in polar coordinates.  In all cases, the $Q$ and $R$ coordinates are left in pixels and conversion to calibrated units was done after machine learning was performed.  All these operations related to disk detection are covered in \href{https://github.com/py4dstem/py4DSTEM_tutorials}{\texttt{py4DSTEM} tutorials}.  After this, the \texttt{pointsarray} can be used in machine learning with \texttt{scikit-learn} functions.  In many cases, specific angular ranges were defined and only diffraction points within these ranges were used.    

\section{Accessible version of the L2 cluster plot for the FeV film}

The L2 cluster plot of Figure 4 had the crystals plotted using a rainbow color scheme to encode the azimuthal orientation of the brightest diffraction spot in each cluster.  Whilst, this makes for a nice plot for many people, those with some form of color vision deficiency would not be able to distinguish many of these colors.  For this reason, an alternate version is made with the crystals colored in a yellow-blue scheme that should be distinguishable for most persons, and this is presented in Fig. \ref{fig:FigS1_accessible}.

\setcounter{figure}{0}
\makeatletter 
\renewcommand{\thefigure}{S\@arabic\c@figure}
\makeatother

\begin{figure}
    \centering
    \includegraphics[width=1\linewidth]{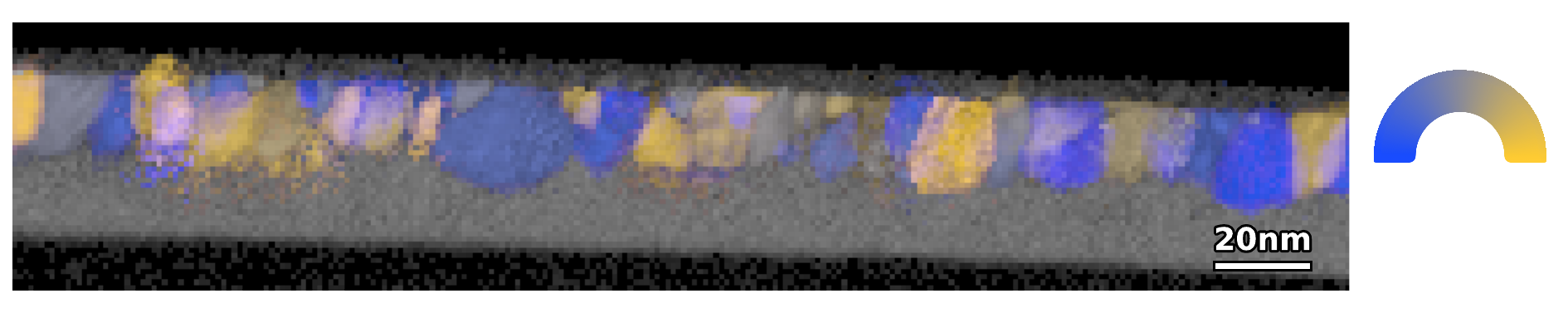}
    \caption{The same plot as shown Figure 4 of the paper, but plotted in a yellow-blue color scheme for persons with color vision deficiency.}
    \label{fig:FigS1_accessible}
\end{figure}

\section{Additional TEM images of plessite}

\begin{figure}
    \centering
    \includegraphics[width=0.5\linewidth]{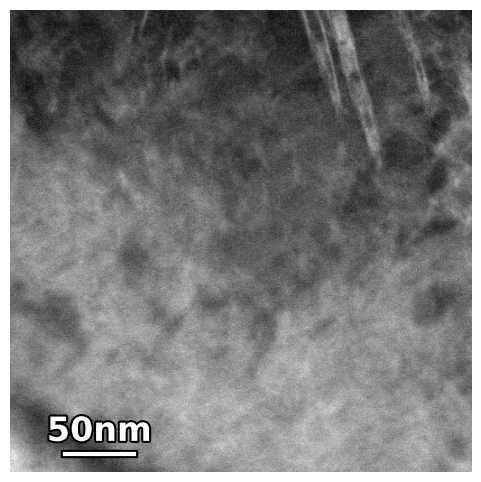}
    \caption{HAADF STEM image partially covering the area also imaged in Figures 6 and 7 of the main paper.  The largest needles are observed but all the finer needles are, at best, only visible as faint disturbances to the contrast.}
    \label{fig:FigS2HAADFplessite}
\end{figure}

The plessite had some fine contrast and the larger needles were visible in TEM and in HAADF STEM (as shown in Fig. \ref{fig:FigS2HAADFplessite}).  However, the rest of the fine scale contrast seen in Figs. 6 and 7 of the main paper was hardly visible and certainly not easy to definitively say anything about.  At best, there are faint dips in contrast in short linear features, but such things are hardly interpretable from a HAADF image.

\begin{figure}
    \centering
    \includegraphics[width=0.5\linewidth]{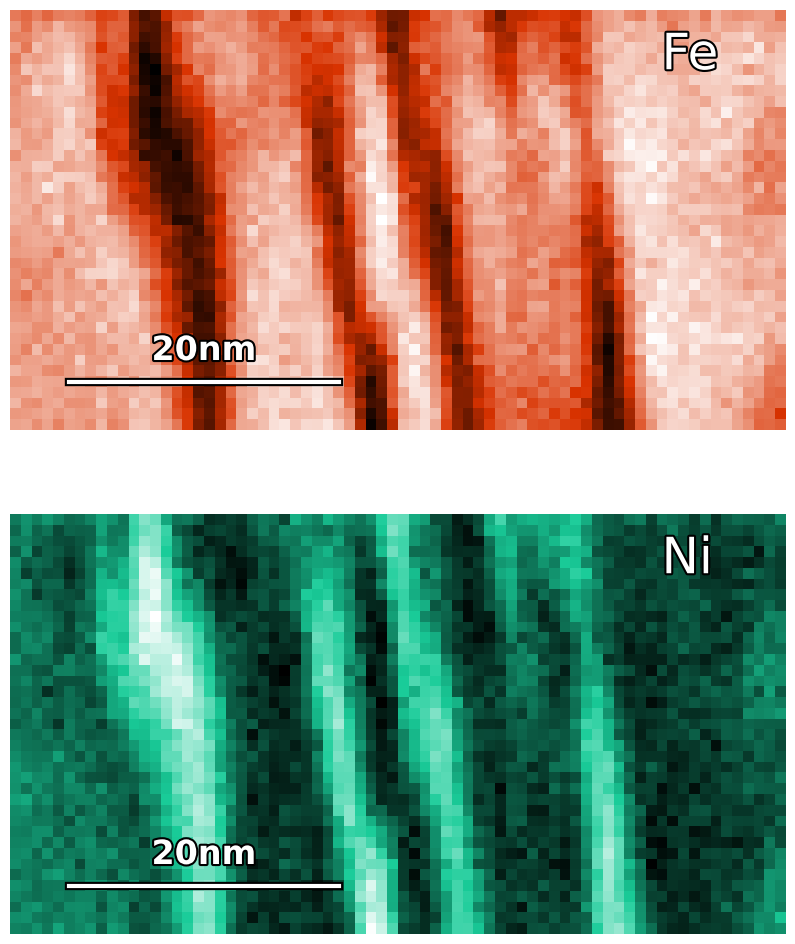}
    \caption{Elemental maps determined from EELS in a box across the two adjacent needles in the top right of Figure \ref{fig:FigS2HAADFplessite}.  Left shows the Fe distribution and right shows the Ni distribution.}
    \label{fig:FigS3EELSmap}
\end{figure}

Elemental mapping with EELS across the larger needles was performed using standard conditions used in many previous studies at the Kelvin Nanocharacterisation Centre using a Gatan GIF Quantum ER as summarized in Table \ref{tab:Table1}.  These were then processed into simple background-subtracted elemental maps using \textit{hyperspy} (\cite{hyperspy}) and made into false color maps with \textit{matplotlib} as shown in Fig. \ref{fig:FigS3EELSmap}.  These clearly show that the edges of these needles are enriched in Ni, but the cores are not, in accordance with the finding from the Clustering analysis that the edges are taenite/tetrataenite and the cores are kamacite.

\begin{figure}
    \centering
    \includegraphics[width=0.5\linewidth]{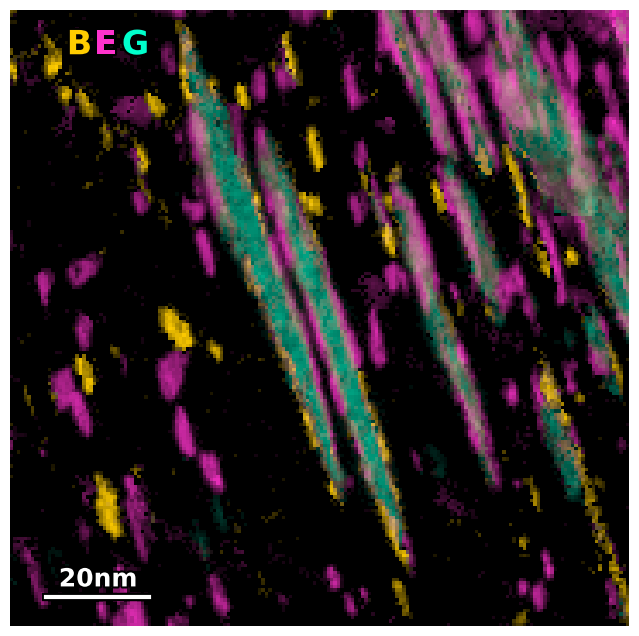}
    \caption{Combining three L2 clusters concerning the largest needles in one image.  This clearly shows that the orientations of taenite/tetrataenite associated with clusters B and E are both found on the rims of the kamacite needles of cluster G.}
    \label{fig:FigS4clusterssuperimposed}
\end{figure}

Combining three of the L2 cluster images from Figure 7 into one is shown in Figure \ref{fig:FigS4clusterssuperimposed} to demonstrate that the rims of the large kamacite needles coincide with some of the features in the taenite/tetrataenite clusters B and E.

\begin{table}
    \centering
    \begin{tabular}{|c|c|}
        \hline
        \textbf{Setting} & \textbf{Value}\\
        \hline
        Accelerating Voltage & 200 kV\\
        Condensor Setting & 6C \\
        Camera Length & 2 cm\\
        Mode & DualEELS\\
        Low loss energy shift & 0 eV\\
        High loss energy shift & 180 eV\\
        Low loss acquisition time & 500 \textmu s\\
        High loss acquisition time & 50 ms\\
        Entrance Aperture & 2.5 mm\\
        Read out mode & Fast 5$\times$1 binning\\
        Drift Correction & Off\\
        \hline
    \end{tabular}
    \caption{Acquisition conditions for EELS maps}
    \label{tab:Table1}
\end{table}

\section{References}
\printbibliography